\documentclass[useAMS]{gGAF2e}

\begin{document}
\doi{10.1080/03091920xxxxxxxxx}
 \issn{1029-0419} \issnp{0309-1929} \jvol{00} \jnum{00} \jyear{2009} 

\markboth{S. Schmitz and A. Tilgner}{Rotating Rayleigh-B\'enard convection}

\title{Transitions in turbulent rotating Rayleigh-B\'enard convection}

\author{S. Schmitz and
A. Tilgner$^\ast$\thanks{$^\ast$Corresponding author. Email:
andreas.tilgner@physik.uni-goettingen.de}
\vspace{6pt}\\\vspace{6pt}  Institute of Geophysics, University of G\"ottingen,
Friedrich-Hund-Platz 1, 37077 G\"ottingen, Germany\\\vspace{6pt}\received{v3.3 released February 2009} }

\maketitle

\begin{abstract}
Numerical simulations of rotating Rayleigh-B\'enard convection are presented for
both no slip and free slip boundaries. The goal is to find a criterion
distinguishing convective flows dominated by the Coriolis force from those
nearly unaffected by rotation. If one uses heat transport as an indicator of
which regime the flow is in, one finds that the transition between the flow
regimes always occurs at the same value of a certain combination of Reynolds, Prandtl
and Ekman numbers for both boundary conditions. If on the other hand one uses
the helicity of the velocity field to identify flows nearly independent of
rotation, one finds the transition at a different location in parameter space.
\bigskip

\bigskip

\end{abstract}

\section{Introduction}

At least two flow regimes exist in rotating Rayleigh-B\'enard convection in
plane layers: Near the onset of convection, the nonlinear advection terms are
small and the Coriolis force dominates the dynamics, provided the Ekman number
is small enough. The controlling effect of the Coriolis force is the
distinguishing feature of the first regime. If the Rayleigh number is increased
at fixed Ekman and Prandtl numbers, the advection term in the Navier-Stokes
equation becomes larger than the Coriolis term (which is linear in velocity) so
that rotation becomes irrelevant and the flow behaves as if rotation was absent.
This defines the second regime. For the purpose of predicting the Nusselt
number, it was found useful in \cite{Schmit09} to introduce a transitional
regime into the classification, in which the Nusselt number obeys a power law
different from those observed in the two other regimes. There is an ongoing
debate concerning the parameters at which the transition between the first and
second regimes occurs (\cite{Rossby69, Aurnou07, Liu09, King09, Schmit09}).
While there is by now ample evidence that the naive
criterion that the Rossby number equals one at the transition is inadequate,
there is no agreement on what the correct criterion is. Recently, it was shown
by \cite{King09} that experimental heat flux data are compatible with the idea
that the flow is in one regime or the other depending on whether the thermal
boundary layer is thicker than the Ekman layer or vice versa. A numerical study
by \cite{Schmit09} avoided Ekman boundary layers by employing stress free
boundary conditions. A transition between the two regimes still occurs and an
analysis of the asymptotic behavior of the Nusselt number leads to a transition
criterion based upon a combination of the Reynolds, Prandtl and Ekman numbers.

The present paper follows up on the numerical simulations presented in
\cite{Schmit09} and intends to answer two questions: First, how important are
the boundary conditions? Is the criterion found in \cite{Schmit09} still
relevant for no slip boundaries? And second, is the classification of a given
flow into the different regimes independent of the quantity used for that
classification? Both \cite{Schmit09} and \cite{King09} classified flows
according to their Nusselt number. Here, we will also look at a quantity derived
from
the velocity field which is important for the dynamo effect, namely the
helicity.

\section{The mathematical model}

A plane layer of thickness $d$, filled with fluid of kinematic
viscosity $\nu$, thermal diffusivity $\kappa$, and thermal expansion coefficient
$\alpha$ rotates with angular velocity $\Omega$ about the
$z-$axis. This axis is perpendicular to the layer.
Gravitational acceleration $g$ is pointing in the negative
$z-$direction. The temperatures of the top and bottom boundaries are fixed at
$T_0$
and $T_0 + \Delta T$, respectively. These two boundaries are assumed to be
either free
slip or no slip, whereas periodic boundary conditions are applied in the $x-$
and
$y-$directions. The equations of evolution are made non-dimensional by using
$d^2/\kappa$, $d$ and $\Delta T$ for units of time, length, and temperature,
respectively. These equations then become within the Boussinesq approximation
for the dimensionless velocity $\bm v(\bm r, t)$ and temperature $T(\bm r, t)$:

\begin{equation}
\partial_t \bm v + (\bm v \cdot \nabla)\bm v +2 \frac{Pr}{Ek} \hat{\bm z} \times
\bm v = -\nabla p + Pr~ \nabla^2 \bm v
+ Ra~ Pr~ T \hat{\bm z}
\label{eq:NS}
\end{equation}

\begin{equation}
\nabla \cdot \bm v =0
\label{eq:div}
\end{equation}

\begin{equation}
\partial_t T + \bm v \cdot \nabla T 
= \nabla^2 T
\label{eq:temp}
\end{equation}
%
$\hat{\bm z}$ is the unit vector in $z-$direction and $p$ collects the pressure
and the centrifugal acceleration. The boundary conditions state in terms of the
adimensional
quantities that
$T(z=0)=1$, $T(z=1)=0$, and free slip boundary conditions require that
$v_z=\partial_z v_x=\partial_z v_y=0$ at both
$z=0$ and $z=1$, whereas $v_x=v_y=v_z=0$ for no slip boundaries.
Three independent dimensionless control parameters appear: The Rayleigh number
$Ra$, the Ekman number $Ek$, and the Prandtl number $Pr$. They are defined by:
\begin{equation}
Ra = \frac{g \alpha \Delta T d^3}{\kappa \nu} ~~~,~~~
Ek = \frac{\nu}{\Omega d^2} ~~~,~~~
Pr = \frac{\nu}{\kappa}.
\label{eq:RaEkPr}
\end{equation}
The Reynolds number $Re$ and the Nusselt number $Nu$ are computed as:
\begin{equation}
Re = \frac{1}{Pr} \sqrt{\frac{1}{V} \int \overline{\bm v^2} dV} ~~~,~~~
Nu = - \frac{1}{A} \int \overline{\partial_z T} dA.
\label{eq:ReNu}
\end{equation}
The overbar denotes average over time and the integrals extend over the
computational volume $V$ for $Re$ and over the surface $A$ of either the top or
the bottom boundary for $Nu$.

The equations of motion were solved with the same spectral method as used in
\cite{Schmit09}. Resolutions reached up to 129
Chebychev polynomials for the discretization of the $z-$coordinate
and $256 \times 256$ Fourier modes in the $(x,y)-$plane.
The periodicity lengths along the $x-$ and $y-$ directions were always chosen to
be identical. Since the typical size of flow structures
varies considerably as a function of the control
parameters in rotating convection, it is not useful to use a single aspect ratio
throughout all simulations, where the aspect ratio is defined as the ratio
of the periodicity length
in the $(x,y)-$plane and the layer height. The
aspect ratio was adjusted for each $Ek$ to fit at least 8 columnar vortices
along both the
$x-$ and $y-$directions at the onset of convection, and kept constant as $Pr$
and $Ra$ were varied.

\begin{figure}
\begin{center}
\includegraphics[width=8cm]{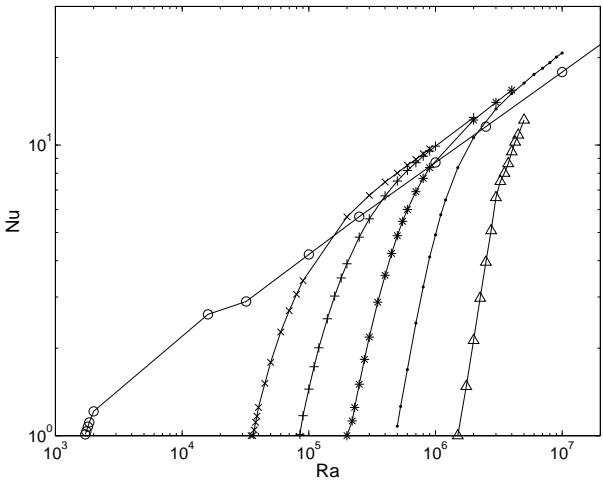}
\caption{$Nu$ as a function of $Ra$ for
no slip boundaries at $Pr=7$ and $Ek=3.4 \times 10^{-3}$ (x),
$Ek=1.7 \times 10^{-3}$ (+), $Ek=8.7 \times 10^{-4}$ (*),
$Ek=4.7 \times 10^{-4}$ ($\bullet$), and $Ek=2 \times 10^{-4}$ ($\Delta$). The open
circles are for zero rotation.}
\label{fig0}
\end{center}
\end{figure}

\begin{figure}
\begin{center}
\includegraphics[width=8cm]{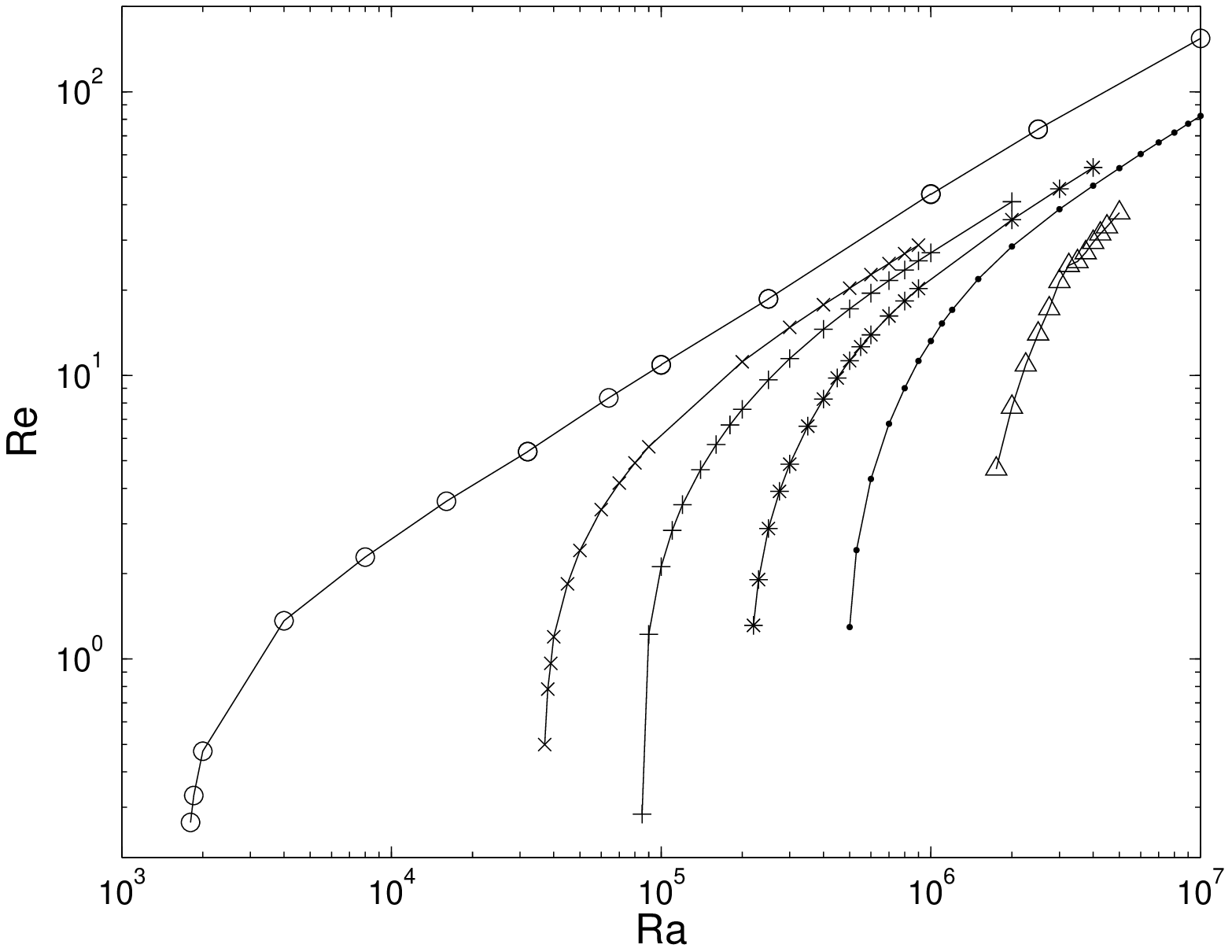}
\caption{$Re$ as a function of $Ra$ for
no slip boundaries at $Pr=7$ and $Ek=3.4 \times 10^{-3}$ (x),
$Ek=1.7 \times 10^{-3}$ (+), $Ek=8.7 \times 10^{-4}$ (*),
$Ek=4.7 \times 10^{-4}$ ($\bullet$), and $Ek=2 \times 10^{-4}$ ($\Delta$). The open
circles are for zero rotation.}
\label{fig0b}
\end{center}
\end{figure}

\begin{figure}
\begin{center}
\includegraphics[width=8cm]{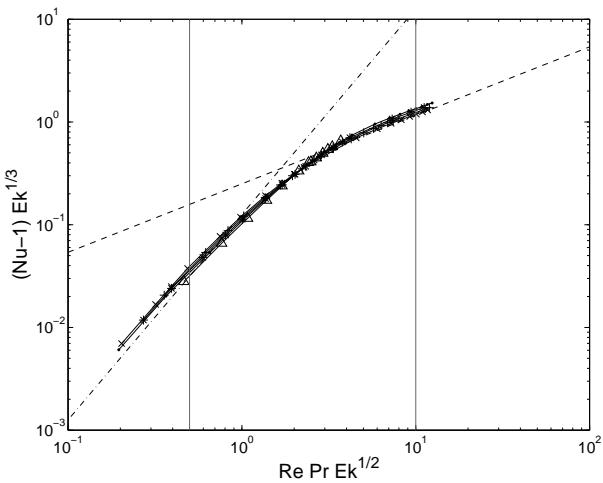}
\caption{$(Nu-1) Ek^{1/3}$ as a function of $Re \, Pr \, Ek^{1/2}$ for 
the same data and with the same symbols as in figs. \ref{fig0} and \ref{fig1}.
The dashed lines are power laws with
exponents 2 and 2/3. The points inside the interval delineated by the two
vertical lines reappear in figure \ref{fig2}.}
\label{fig1}
\end{center}
\end{figure}

\begin{figure}
\begin{center}
\includegraphics[width=8cm]{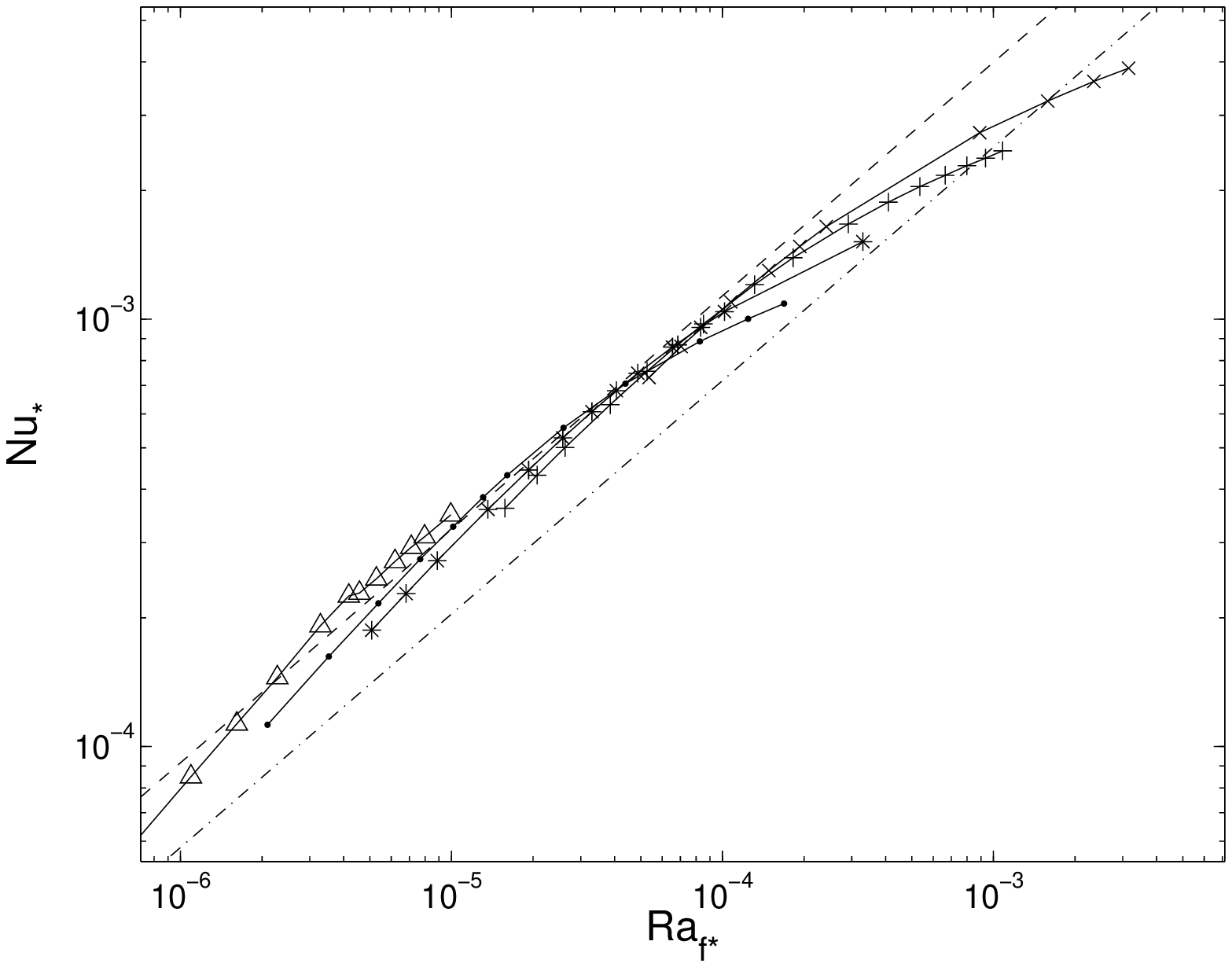}
\caption{$Nu_*$ as a function of $Ra_{f*}$. The symbols have the same meaning as
in
fig. \ref{fig0}. This figure contains only those data points which lie in the
interval marked by vertical lines in fig. \ref{fig1}. The straight lines
indicate the functions
$Nu_*=0.11 \cdot Ra_{f*}^{0.55}$ (lower, dot dashed line) which fits well to
simulations with free slip boundaries, and
$Nu_*=0.17 \cdot Ra_{f*}^{0.55}$ (upper, dashed line).}
\label{fig2}
\end{center}
\end{figure}

\section{No slip boundaries \label{section_no_slip}}

Computations for no slip boundaries are much more demanding than in the free
slip case. That's why the parameter range covered by the data in figure
\ref{fig0} is smaller than in the equivalent figure in \cite{Schmit09}.
Nonetheless, the data suffice to show that the behavior is essentially the same
for both types of boundary conditions. Figure \ref{fig0} shows $Nu(Ra)$ for no
slip boundary conditions. One recognizes the well known pattern of a delayed
onset of convection followed by a steep rise.
$Nu$ in the rotating flow may even exceed its value for zero rotation for no
slip boundary conditions (\cite{Zhong09}). The delayed onset of convection is also
visible in the graph of Reynolds number versus Rayleigh number (figure \ref{fig0b}).

Figure \ref{fig1} uses a data reduction motivated by the results for free slip
boundary conditions . Some of the arguments of \cite{Schmit09} are recapitulated
here to make the paper self-contained.
Low values of ${Re} \, {Pr} \, {Ek}^{1/2}$ 
correspond to laminar flows
near the onset of convection. Forming the dot product of eq. (\ref{eq:NS}) and
$\bm v$, integrating over the whole volume and averaging over time, one finds
\begin{equation}
\epsilon = ({Nu}-1){Ra}
\label{eq:epsilon}
\end{equation}
where $\epsilon=\frac{1}{V} \int <(\partial_i v_j) (\partial_i v_j)> dV$ is the
adimensional
average dissipation rate of kinetic energy. In a laminar flow, one expects
$\epsilon \propto ({Re}\, {Pr})^2/\lambda^2$, where $\lambda$ is a
characteristic
length scale of the flow. For ${Pr} > 0.676$, convection starts at a
critical
Rayleigh number ${Ra}_c$ obeying ${Ra}_c \propto
{Ek}^{-4/3}$ and forms stationary cells of size
$\lambda_c$ with $\lambda_c \propto {Ek}^{1/3}$ (\cite{Chandr61}). Eq.
(\ref{eq:epsilon}) becomes ${Ek}^{-2/3}{Re}^2 {Pr}^2
\propto ({Nu}-1) {Ra}$. Close to
onset, ${Ra} \approx {Ra}_c$ and therefore $({Nu}-1)
\propto {Re}^2 {Pr}^2 {Ek}^{2/3}$. 
For large values of ${Re} \, {Pr} \, {Ek}^{1/2}$, one
should approach the non-rotating case and find a law independent of 
${Ek}$. There is no reliable theory for convection far from the onset so
that we have to rely on numerical results. Simulations show that 
$({Nu}-1) \propto ({Re} \, {Pr} \,)^{2/3}$ for zero
rotation. Both asymptotes, near onset and far from it, become straight lines in
a logarithmic plot of
$({Nu}-1) {Ek}^{1/3}$ vs. ${Re} \, {Pr} \, {Ek}^{1/2}$.

Figure \ref{fig1} shows 
$(Nu-1) Ek^{1/3}$ as a function of $Re \, Pr \, Ek^{1/2}$ for no slip boundary
conditions and should be compared with figure 2 of \cite{Schmit09}. The dashed
lines have the slopes of the asymptotes which $(Nu-1) Ek^{1/3}$ has to follow
either in the limit of large Rayleigh numbers (when rotation plays no role) or
near the onset of convection. The prefactors in these power laws are
obtained from best fits to the simulations with free slip boundary conditions of
\cite{Schmit09} so that a direct comparison is possible with the results of
computations with no slip boundary conditions shown by points in the figure. It
is seen that both boundary conditions have asymptotes with the same exponents.
Most importantly, for both boundary conditions, the crossing of the two
asymptotes near $Re \, Pr \, Ek^{1/2}=2$ separates the convection
dominated by rotation from the convection virtually unaffected by rotation.

It is noteworthy that in the no slip case of figure \ref{fig1}, the asymptotes
provide us with reasonably good fits throughout the entire parameter range. In
the case of free slip boundary conditions in \cite{Schmit09}, the interval 
$1/2 < Re \, Pr \, Ek^{1/2} < 10$ (indicated by vertical bars in figure
\ref{fig1}) required special treatment. It turned out to be useful to introduce
nondimensional parameters independent of molecular diffusivities already
discussed in previous work (\cite{Christ02, Aurnou07}): 
$Nu_*= Nu \, Ek/Pr = Q/(\rho c_p \Delta T \Omega d)$ and
$Ra_{f*}=Nu_* \, Ra \, Ek^2/Pr = (g \alpha Q) / (\rho c_p \Omega^3 d^2)$, where
$\rho$ denotes density and $c_p$ heat capacity. An envelope to the data in the
interval $1/2 < Re \, Pr \, Ek^{1/2} < 10$ is given by 
$Nu_*=0.11 \cdot Ra_{f*}^{0.55}$ for free slip boundaries. Figure \ref{fig2}
shows that a similar conclusion holds for no slip boundaries after a change
in the prefactor: $Nu_*=0.17 \cdot Ra_{f*}^{0.55}$ is more appropriate in this
case.

\section{Free slip boundaries: Helicity \label{section_free_slip}}

\begin{figure}
\begin{center}
\includegraphics[width=8cm]{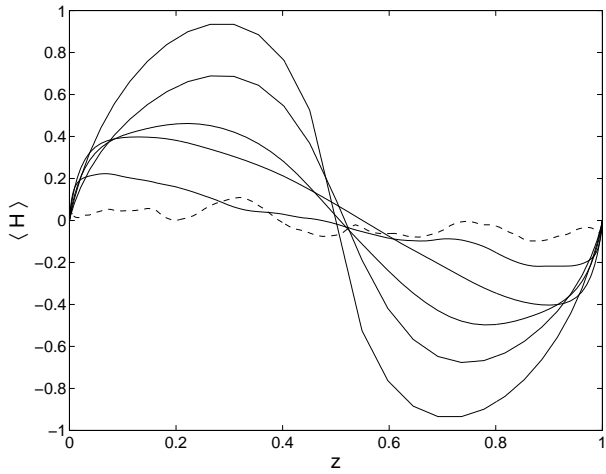}
\caption{Average helicity $\langle H \rangle$ as a function of height $z$ for 
no slip boundaries, $Ek=2 \times
10^{-3}$ and $Pr=7$. In going from the curve drawn with a continuous line of
largest amplitude to the one with smallest amplitude, the Rayleigh number is
$Ra= 1.0 \times 10^5$, $2.0 \times 10^5$, $5.0 \times 10^5$, $1.0 \times 10^6$,
and $1.0 \times 10^7$. The dashed curve is for $Ra=1.0 \times 10^8$.}
\label{fig3}
\end{center}
\end{figure}

In order to quantify the influence of rotation on the flow structure, we
investigate the helicity $H$, or more precisely the correlation between
vorticity and velocity, defined by:
\begin{equation}
H=\frac{(\nabla \times \bm v) \cdot \bm v}{|\nabla \times \bm v| \, |\bm v|}.
\end{equation}
This correlation is now averaged over horizontal planes, and ideally over time.
However, helicity was not saved during the production runs so that it had to be
deduced from single snapshots. Assuming ergodicity of the flow, the spatial
average replaces to some extent the temporal average. The noise in the figures
below is low enough so that further simulations did not seem warranted. The
average helicity $\langle H \rangle$, defined by 
\begin{equation}
\langle H \rangle =\frac{1}{A} \int dx \int dy \, H
\end{equation}
where $A$ is the area of a horizontal plane, is shown in figure \ref{fig3}. As
expected (\cite{Chandr61}), rotation introduces helicity of one sign in one half of layer and of
the opposite sign in the other half ($z=0.5$ is not an exact plane of symmetry
in figure \ref{fig3} because of the missing time average). As $Ra$ increases,
the advection term increases compared with the Coriolis term and the helicity
decreases. The maximum of $\langle H \rangle$ is reached at a point at a smaller and smaller
distance from the boundaries as $Ra$ is increased. Helicity is induced by the presence of
the boundaries because they force in- and outflow out of departing or arriving
plumes which is spun up by the Coriolis force. When turbulence tends to destroy
the correlation between vorticity and velocity present in helical structures, it
is easiest to do so away from the walls.

A suitable global measure for the influence of the rotation on the velocity
field is the rms fluctuation in $\langle H \rangle (z)$, i.e.
\begin{equation}
H_{rms}=\int_0^1 \langle H \rangle ^2 dz.
\end{equation}

\begin{figure}
\begin{center}
\includegraphics[width=8cm]{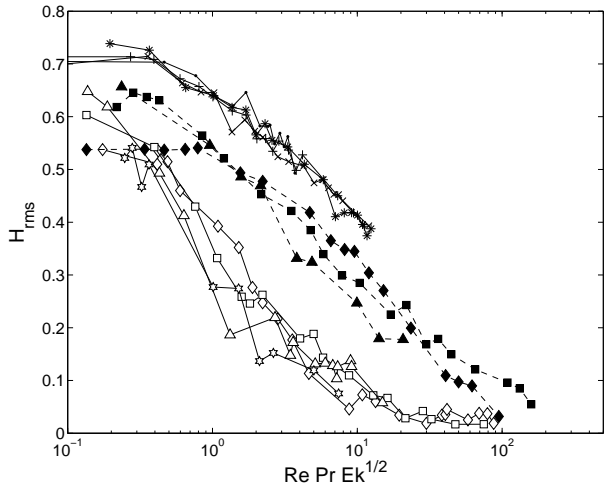}
\caption{$H_{rms}$ as a function of $Re \, Pr \, Ek^{1/2}$. The filled symbols
are for $Pr=7$ and the empty symbols for $Pr=0.7$. The Ekman numbers are 
$Ek=2.0 \times 10^{-2}$ (diamonds), $2.0 \times 10^{-3}$ (squares), $2.0 \times
10^{-4}$ (triangles) and $2.0 \times 10^{-5}$ (stars). The remaining symbols indicate
no slip boundaries and $Ek=1.7 \times 10^{-3}$ (x), $Ek=8.7 \times 10^{-4}$ (+),
$Ek=4.7 \times 10^{-4}$ (*), and $Ek=2 \times 10^{-4}$ ($\bullet$), all for $Pr=7$.}
\label{fig4}
\end{center}
\end{figure}

\begin{figure}
\begin{center}
\includegraphics[width=8cm]{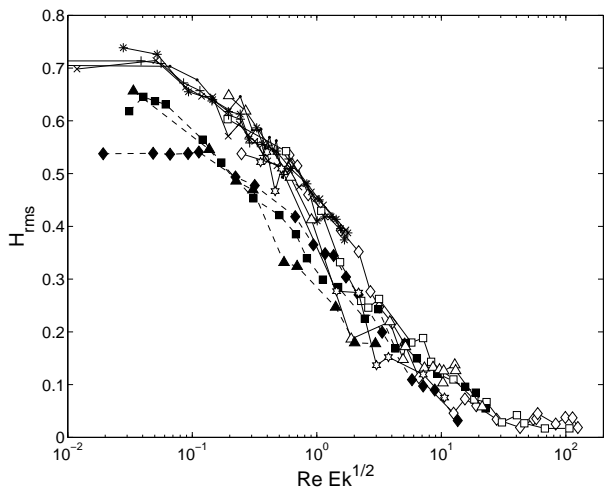}
\caption{$H_{rms}$ as a function of $Re \, Ek^{1/2}$. The symbols are the same
as in figure \ref{fig4}.}
\label{fig5}
\end{center}
\end{figure}

Figure \ref{fig4} gives $H_{rms}$ as a function of $Re \, Pr \, Ek^{1/2}$.
There is a reasonable collapse of the data for a given $Pr$, which shows that
the combination $Re \, Pr \, Ek^{1/2}$ captures the right $Re$ and $Ek$
dependences, but the points for different $Pr$ do not fall on top of each other.
However, if one plots $H_{rms}$ as a function of $Re \, Ek^{1/2}$, one again
obtains a good collapse of the data points, at least at large $Re \, Ek^{1/2}$
(see figure \ref{fig5}). Within the
noise on $H_{rms}$, the flow is indistinguishable from nonrotating
convection for $Re \, Ek^{1/2} > 50$. The transition criteria are thus not
identical when based on the Nusselt number or on the helicity. The transition
occurs at a certain value of $Re \, Pr \, Ek^{1/2}$ in as far as heat flux is
concerned, and at a certain value of $Re \, Ek^{1/2}$ for helicity.

\section{Discussion and conclusion}

It was shown in section \ref{section_no_slip} that according to the Nusselt
number, $Re \, Pr \, Ek^{1/2} = 2$ separates flows dominated by the Coriolis
force from flows unaffected by rotation for both types of boundary conditions,
free slip and no slip. The combination $Re \, Pr \, Ek^{1/2}$ can be recognized
as the Peclet number based on the thickness of the Ekman layer. This
interpretation is not directly useful, however. It is not clear why this Peclet
number should matter for the heat flux, because the heat flux is crossing
diffusively the velocity boundary layer in which velocity and temperature
gradient are mostly perpendicular to each other so that $\bm v \cdot \nabla T = 0$ and
$\bm v$ does not enter the temperature equation (\ref{eq:temp}) within the boundary layers. 
In addition, there is no Ekman layer for free slip
boundaries, and yet, the transition occurs at the same values of the control
parameters. 

One may argue that a length scaling as $Ek^{1/2}$ is not completely foreign to
free slip boundaries, either. \cite{Hide64} calculates a velocity boundary layer
thickness in $Ek^{1/2}$ near a free slip boundary provided that the density of
the fluid varies along the surface. This cannot be the case within the
Boussinesq approximation on an isothermal boundary, so that this mechanism
cannot create a length in $Ek^{1/2}$ in our simulations. \cite{Julien96} also
find an $Ek^{1/2}$ layer in the velocity field if the thermal boundary layer
thickness varies laterally. Their calculation posits a lateral temperature
variation which is not advected by the horizontal velocity, so that this result
is not directly applicable to self-consistent simulations.
Nonetheless, these analytical calculations are motivation enough to scan the
velocity profiles obtained from the numerical simulations for a length scaling
as rapidly as $Ek^{1/2}$ as a function of the Ekman number. However, no such
length could be found in the case of free slip boundaries.

As section \ref{section_free_slip} has shown, the transition from a flow
dominated
by rotation to a flow unaffected by rotation is not well defined, anyway. The
transition occurs at different values of the control parameters depending on
whether the classification is based on the Nusselt number or on the helicity of
the velocity field. The helicity is down to negligible magnitude for
$Re \, Ek^{1/2} > 50$. We will now apply this result to the Earth's core. Most
data in this paper were obtained for free slip boundaries, but the close
agreement between free slip and no slip boundaries found here encourages us to
apply the transition criteria to the Earth's core nonetheless.
For the Earth's core, the generally accepted values
\footnote{There is an error in one numerical value in \cite{Schmit09}: The value
of $Re \, Pr \, Ek^{1/2}$ for the core parameters quoted there should be 13, not 5.}
of $\nu= 5 \times 10^{-7} m^2/s$, $\Omega=7.29 \times 10^{-5} s^{-1}$ and a typical
flow velocity of $5 \times 10^{-4} m/s$ lead to $Re \, Ek^{1/2} = 82$. This
estimate casts doubts on the picture of a geodynamo operating as an
$\alpha^2-$dynamo with an $\alpha-$effect due to the helicity of the flow.

This work was supported by the Deutsche Forschungsgemeinschaft (DFG).


\end{document}